\documentclass[10pt]{iopart}
\usepackage{graphicx,iopams}
\usepackage[normalem]{ulem}

\usepackage{color}

\begin{document}

\title{Ising model with variable spin/agent strengths}

\author{M.~Krasnytska$^{1,2}$,  B. Berche$^{3,2}$, Yu.~Holovatch$^{1,2,4}$,  R.~Kenna$^{4,2}$}

\address{$^{1}$ Institute for Condensed Matter Physics of the National Acad. Sci. of Ukraine,
             79011 Lviv, Ukraine}

\address{$^{2}$ $\mathbb{L}^4$ Collaboration \& Doctoral College for the
Statistical Physics of Complex Systems,
Leipzig-Lorraine-Lviv-Coventry}

\address{$^{3}$  Laboratoire de Physique et Chimie Th\'eoriques, Universit\'e de
Lorraine, BP 70239, 54506 Vand\oe uvre-les-Nancy Cedex, France}

\address{$^{4}$ Centre for Fluid and Complex Systems, Coventry University, Coventry,
CV1 5FB, United Kingdom}
 
\ead{submissions@iop.org}
\vspace{10pt}
\begin{indented}
\item[]June 6, 2020
\end{indented} 
 

\begin{abstract}
We introduce varying spin strengths to the Ising model, a
central pillar of statistical physics. With inhomogeneous physical
systems in mind, but also anticipating interdisciplinary
applications, we present the model on  network structures of varying
degrees of complexity. 
This allows us explore the interplay of two types of randomness: individual strengths of spins or agents and collective  connectivity between them.
We solve {the model}  for the generic case of
power-law spin strength and find that, with a self-averaging free
energy, it has a rich phase diagram with new universality
classes. Indeed, the degree of complexity added by quenched variable spins is
on a par to that added by endowing simple networks with increasingly
realistic geometries. The model is suitable for investigating emergent phenomena
in many-body systems in contexts where non-identicality of spins or
agents plays an essential role  and for exporting statistical
physics concepts beyond physics. 
\end{abstract}

%
%
\submitto{Journal of Physics: Complexity}
%
%
\ioptwocol

It is difficult to overestimate the importance of the Ising model
(IM) in  physics  and beyond. It was invented 100 years ago by W.
Lenz~\cite{Lenz20} and  solved in one dimension with
nearest-neighbour interactions by E. Ising~\cite{Ising25}. It draws
on works of R. Kirwan and W. Weber~\cite{Kirwan,Weber} who proposed
that vanishing  magnetization of  macroscopic para- and
ferromagnetic bodies originates from random disordering of identical
elementary magnets (spins) at microscopic levels. While their ideas
also explained magnetic saturation in strong external fields, they
failed to explain gradual response to  weak ones.
 E.A. Ewing~\cite{Ewing}  introduced interactions to mend this shortcoming and P. Weiss \cite{Weiss} used the idea in the mean-field approximation.
The IM itself sits between the Kirwan-Weber non-interacting picture
and the Curie-Weiss maximally  interacting one and manifests
spontaneous spin alignment as well. Yet it is its simplicity,
coupled with the phenomenon of universality, that lies behind its
applicability to many real-world many-body  phenomena in physics and
beyond~\cite{Ising17}.

More  elaborate models  describe specific physical phenomena with
greater precision, using greater levels of sophistication either in
spin variables themselves or  interactions among them. For the
former case, the simple polarity,  wherein  spins have two possible
states, is maintained in the  continuous spin Ising
model~\cite{vanBeijeren78,Bayong99}. This is relaxed in the Potts
model while maintaining spin discreetness \cite{Potts52,Wu82} and
the $m$-vector model goes further by introducing vector variables
$\vec{\sigma}_i$~\cite{Stanley68,Stanley71}. Besides loosening Ising
constraints on spins, they can be relaxed for interactions as well
and simple extensions include next-nearest {neighbour} interactions,
equivalent neighbours \cite{Luijten96} or probabilistic long-range
interactions~\cite{Fisher92}, disorder and
frustrations~\cite{spin_glasses}.

These variants deliver new universality classes, and, while
associated models  have been  successful in their respective
contexts, they  rest on the same  concept of interacting  entities
being identical. But not all particles are identical --- they may
have different  inhomogeneities in  internal degrees of freedom
manifestating, e.g., different magnetic moments. None of the above
models deal with this feature.

The IM  brings physics beyond its traditional realms too. E.g., it
models  phenomena as diverse as cancer cells' response to
chemotherapy~\cite{Cancer}; yield patterns in  trees~\cite{trees};
and advertising in duopoly markets~\cite{Sznajd}. 
{Although each of these involves interactions between entities that are not spins, the IM is well placed to describe their collective properties. 
The importance of interactions in systems beyond physics is now well understood in society at large, as are consequences of not understanding this~{\cite{commonletter1,commonletter2}.}}
Still, a protest often encountered when {seeking to export} statistical concepts   is ``people are not
atoms''~\cite{Stauffer}. 
{The term ``agent'' was introduced to reflect this linguistically but it does automatically de-binarise the interacting entities.}
Assurances such as ``{it is} the law of big numbers {which} allows the application of statistical physics methods''~\cite{Stauffer}
are often not understood, welcomed or accepted.
Addressing these  issues {(the individuality of agents as well as the role of interactions)} becomes important if we are to communicate
physics to interdisciplinary colleagues or authorities; not all
cells, trees or bankers are the same and there
are different degrees of contagiousness among infected
agents.

{This} suggest a new variant to the {IM} which, besides accounting for
particle peculiarities,  addresses  non-identical agents. We
introduce the model on network topologies for reasons applicable to
both scenarios. In physics, there are nanosystems with topologies
more akin to networks  than lattices~\cite{nanosystems}. Varying
spin strength in such systems may serve to model polydispersity in
elementary magnetic moments~\cite{Russier12}. Furthermore,
 perfect lattice structures are not common beyond physics and never encountered in sociophysics.
Another example may be taken from neuroscience; 
in their recent review, Lynn and Bassett invite ``creative effort'' to ``tackle some of the most pressing questions surrounding the inner workings of the mind''~\cite{Lynn19}. 
It is well known that cognition rests on heterogeneous connectivity structures in the brain and that neural communication strongly relies on a complex interplay between its network geometry and topology~\cite{Senguin18}.  
Introduction of variability to nodal strengths may likewise contribute to network modeling~\cite{Betzel17}, accounting for differences in how nodes are defined as collections of brain tissue~\cite{Stanley13}. 
In particular, as an interacting unit in a brain network, the strength of a node may reflect its internal structure on a hierarchical level. 
A node may itself be composed of smaller nodes~\cite{Park13} or the changes in node strength may correspond to changes in node functionality over a span of time~\cite{Bagarinao19}.

Here, we suggest and exactly solve an IM with variable spins/agents
on networks of different degrees of complexity. Competition between
individual agent strengths and collective  connectivity generates a
rich phase diagram suitable for the analysis of new universal
emergent phenomena in many-agent systems~\cite{Holovatch17}.

\section*{Methods and Results}

In the presence of a homogeneous  field $H$, the Hamiltonian of the
{IM} reads:
\begin{equation}\label{1}
{\cal H} = - \sum_{i<j} J_{ij} \sigma_i \sigma_j - H \sum_i \sigma_i
\, ,
\end{equation}
For the standard  version spins  $\sigma_i$  take values $\pm 1$, $i=1\dots N$ 
and
coupling $J_{ij}$ is  $1$ for the nearest {neighbour} sites and $0$
otherwise. In sociophysics individuals are  represented as nodes
with binary spins representing different social states. This
limitation is precisely because of the  duality of spins in the
generic IM. It may well be convincing for  ``for'' or ``against''
options in referendums~\cite{Galam} but not all societal activities
are binary and  our model introduces  gradation for individual node
features.

 We endow the spins $\sigma_i$ with  ``strengths'' which vary from
site to site through a  random variable $|\sigma_i|\equiv
\mathcal{S}_i$ drawn from   probability distribution function
$q(\mathcal{S}_i)$. We chose a power-law decay:
\begin{equation}\label{2}
q(\mathcal{S})=c_{\mu}\mathcal{S}^{-\mu}, \hspace{1cm}
\mathcal{S}_{\rm min}  \leq \mathcal{S}  \leq \mathcal{S}_{\rm
max}\, ,
\end{equation}
where $c_{\mu}$ is a normalization constant, with  $\mu>2$  ensuring
finite mean strength $\overline{\mathcal{S}}$ when $\mathcal{S}_{\rm
max}\to \infty$. 
The spin length being quenched random variable,
a proper definition of thermodynamic functions involves configurational
averaging besides the usual Gibbs averaging \cite{Brout59}.\footnote{Since the lengths are quenched variables, the model introduced here is distinct from the continuous spin model.
It should also not be confused with  spin-glass or annealed models for which a large body of literature exists.
The term annealed in this presentation refers to the random graph on which the new model is solved. }

There are several obvious motivations for choosing a power-law form
in the first instance. We don't expect to encounter physical
phenomena with spin strengths  distributed precisely in this manner,
but it is an  established tradition in physics  to take idealised
models to investigate the fundamentals of a concept. Indeed Lenz and
Ising applied this strategy when introducing the model in the first
place. To take examples beyond physics, variable agent strength can
be used to represent degrees of contagiousness in
pandemics~\cite{Covid} or opinion in social systems.

Because of the interactive nature of the {IM}, these strengths
impact on nodes with which a given node interacts --- the greater
the value of $\mathcal{S}_i$ the more contagious or persuasive node
$i$ is. This new element is  closer to real social networks and more
likely to be accepted  beyond physics~\cite{leaders}. Thus the
introduction of variable spins to an established model opens  new
avenues to physics, interdisciplinarity and communication of same.
As we shall see, these  deliver very rich critical behavior and an
onset of new accompanying  phenomena which are of fundamental
interest in their own right.

Since we are exploring new avenues, we consider three graph
architectures which permit exact solutions:
 the complete graph,  the Erd\"os-R\'enyi  graph and scale-free networks.
As in the  Weiss  model, every  nodal pair $\{i,j\}$, is linked in
the first case. The probability of connectedness is also the same
for every nodal pair $p=p_{i,j} = c <1$ in the second case but it
differs in that not every  pair is linked. For the third case, the
node degree distribution   is governed by a power-law decay
\cite{networks,Dorogovtsev08}:
\begin{equation}\label{3}
p(K)=c_{\lambda} K^{-\lambda}, \hspace{1cm} K_{\rm min} \leq  K \leq
K_{\rm max}\, ,
\end{equation}
 for constant  $c_{\lambda}$  and $\lambda>2$.
The adjacency matrix  for an annealed random graph is
\cite{Lee09,Bianconi12,Krasnytska16}:
\begin{eqnarray}\label{4}
 J_{ij}=\left\{
\begin{array}{ccc}
                 1,& \hspace{0.5em} p_{ij}\, ,\\
                 0,& \hspace{0.5em} 1- p_{ij}\, ,
              \end{array}
  \right.
\end{eqnarray}
where $p_{ij}$ is the edge probability  \ any pair  $i$ and $j$. For
$N$ nodes, one assigns a random  degree  $k_i$ to each, taken from
the distribution
\begin{equation}\label{5}
p_{ij}= \frac{k_ik_j}{N\overline{k}}+O(1/N^2)\, ,
\end{equation}
with $\overline{k} =\frac{1}{N}\sum_lk_l$. The expected value of the
node degree is $\mathbb{E} K_i=\sum_j p_{ij}=k_i$ and its
distribution is given by $p(K)$ \cite{Bianconi12,Krasnytska16}.
 The limiting cases $p_{ij}=1$ and $p_{ij}=c$ recover the complete  and  Erd\"os-R\'enyi random graph respectively.

Boltzmann averaging  for the partition function  is bond and
spin-strength   $\{ \mathcal{S} \}$ configuration dependent. For
annealed networks,  averaging over links
\begin{equation}\label{6}
\langle (\dots) \rangle_{\{J\} } = \prod_{i<j}  \Big [
(\dots)_{J_{ij}=1} p_{ij} + (\dots)_{J_{ij}=0} (1-p_{ij}) \Big ]  \,
\end{equation}
delivers equilibrium and  is applied to the partition function vis.:
\begin{equation}
\langle {\cal Z}(\{J\},\{\mathcal{S}\}) \rangle_{\{J\} } = {\cal
Z}(\{k\},\{\mathcal{S}\})\, .
\end{equation}
The corresponding quenched free energy $f(\{k\},\{\mathcal{S}\})$
depends on the  fixed random variables  $\{k_1,k_2,\dots k_N\}\equiv
\{k\}$ and $\{\mathcal{S}_1,\mathcal{S}_2,\dots
\mathcal{S}_N\}\equiv \{\mathcal{S}\}$, {These sequences are taken
as fixed, quenched ones,} so the  free energy $f$ is  obtained by
averaging over them \cite{Brout59}. As we  show below, the
corresponding partition function,  and   thermodynamic functions are
self-averaging;  they do not depend on a particular choice of
$\{k\}$ and $\{\mathcal{S}\}$.

Since the spin product in (\ref{1}) can attain only two values, we
use  the equality
$$
\phi(A\epsilon)=\frac{1}{2}[\phi(A)+\phi(-A)]+\frac{\epsilon}{2}[\phi(A)-\phi(-A)],
\, \epsilon = \pm 1,
$$
to get for the configuration-dependent partition function:
\begin{eqnarray}\label{7}
{\cal Z}(\{k\},\{\mathcal{S}\})=\prod_{i<j}c_{ij}{\rm Sp}_{\sigma}
\Big(
 e^{\sum_{i<j}d_{ij} \sigma_i\sigma_j+\beta H\sum_i \mathcal{S}_i\sigma_i}\Big)\ ,
\\ \label{8}
c_{ij}=\sqrt{a_{ij}^2-b_{ij}^2}, \, d_{ij}=\ln
\frac{a_{ij}+b_{ij}}{a_{ij}-b_{ij}}\, .
  \end{eqnarray}
Here,  {$\beta=T^{-1}$} is {the inverse} temperature,  the trace is  taken
over all spins and we keep $\sigma_i=\pm 1$ representing each spin
value as $\sigma_i\mathcal{S}_i$.
 The coefficients~(\ref{8}) implicitly
depend on $p_{ij}$ and $\mathcal{S}_i$ via
\begin{equation}\label{9}
a_{ij}=1-p_{ij}+p_{ij}\cosh(\beta J\mathcal{S}_i\mathcal{S}_j), \,
b_{ij}=p_{ij}\sinh(\beta J\mathcal{S}_i\mathcal{S}_j).
  \end{equation}

{\bf{The complete graph}} has  $p_{ij}=1$,  from which $c_{ij}=1$,
$d_{ij}=2\beta \mathcal{S}_i\mathcal{S}_j$, and averaging over
spins gives:
\begin{equation}\label{10}
{\cal Z}(\{\mathcal{S}\})=\int_{0}^{+\infty} e^{\frac{-x^2 T
}{2}}\Big[e^{I^+_\mu(\frac{x}{\sqrt{N}})}+ e^{I^-_\mu(\frac{x}{\sqrt{N}})}\Big] dx \, ,
\end{equation}
having used (\ref{2}) with $\mathcal{S}_{\rm max}\to \infty$. We
obtain for small magnetic fields
\begin{equation}\label{11}
I^\pm_\mu(\varepsilon)= N \Big[c_\mu
\varepsilon^{\mu-1}I_\mu(\varepsilon)\pm \frac{
\overline{\mathcal{S}^2}}{T} \varepsilon H \Big],
\end{equation}
with
\begin{equation}\label{12}
I_\mu(\varepsilon)=\int_{\varepsilon}^{\infty}  dz\,
\frac{1}{z^\mu}\ln \cosh z,
\end{equation}
and $\varepsilon=\frac{x}{\sqrt{N}}$. In (\ref{10}) and in all
counterpart integral representations below, we omit  irrelevant
prefactors and numerical coefficients of associated response
functions are presented elsewhere~\cite{Krasnytska20}.

\begin{table*}
\begin{center}
\tabcolsep1.2mm
\begin{tabular}{lcccccc}
\hline \hline  & $\alpha$ & $\alpha_c$ & $\gamma$ & $\gamma_c$ &  $\beta$ &$\delta$  \\
\hline
Line 4 ($\lambda=\mu$) & $\frac{\lambda-5}{\lambda-3} $   & $\frac{\lambda-5}{\lambda-2}$ & 1 &   $\frac{\lambda-3}{\lambda-2}$ &  $ \frac{1}{\lambda-3}$ &  $\lambda-2 $  \\
Region III & $\frac{\lambda-5}{\lambda-3} $   & $\frac{\lambda-5}{\lambda-2}$ & 1  &  $\frac{\lambda-3}{\lambda-2}$ &  $ \frac{1}{\lambda-3}$ &  $\lambda-2 $  \\
Region IV & $\frac{\mu-5}{\mu-3} $   & $\frac{\mu-5}{\mu-2}$ &  1 &  $\frac{\mu-3}{\mu-2}$ &  $ \frac{1}{\mu-3}$ &  $\mu-2 $  \\
Region V, Lines 5,6, Point B & 0    & 0 &  1  &  2/3 &   1/2 &  3  \\
\hline \hline
\end{tabular}
\end{center}
\caption{Critical indices governing temperature and field
dependencies of the specific heat,  susceptibility, and order
parameter in different parts of the phase diagram of Fig.
\ref{fig1}. \label{tab1}}
\end{table*}

\begin{table*}
    \begin{center}
        \tabcolsep1mm
            \begin{tabular}{lcccccc}
                \hline \hline  & $\hat{\alpha}$ & $\hat{\alpha_c}$ & $\hat{\gamma}$
                & $\hat{\gamma_c}$ &
                $\hat{\beta}$ &$\hat{\delta}$ \\
                \hline
                Line 4 ($\lambda=\mu$) & $-\frac{3}{\lambda-2} $   & $-\frac{3}{\lambda-2}$ & 0 &   $-\frac{\lambda-3}{2(\lambda-2)}$ &  $ -\frac{1}{\lambda-3}$ &  $ -\frac{1}{\lambda-2}$   \\
                Point B & $-2$  & $-2$ &  $0$ &  $-2/3$ &   $-1$ &  $-2/3$ \\
                \small{Lines 5, 6} & $-1$  & $-1$ &  $0$ &  $-1/3$ &   $-1/2$ &  $-1/3$  \\
                \hline \hline
            \end{tabular}
            \end{center}
    \caption{Logarithmic-correction exponents in different regions of Fig. \ref{fig1}.
                        Exponents along line 4 ($\mu=\lambda$) and at point~B represent new universality classes.
        \label{tab2}}
    \end{table*}

The partition function (\ref{10})  is independent of
$\{k\},\{\mathcal{S}\}$; for the random configuration $\{k\}$ this
is obvious since $p_{ij}=1$ and for the random spin strength
configuration $\{\mathcal{S}\}$  it is due to self-averaging. This
is a generic feature of the model  as we shall see below. With the
asymptotic behavior of $I_\mu(\varepsilon)$ (\ref{12}) to
hand~\cite{asymptoticsa,asymptoticsb,Krasnytska20} one evaluates
(\ref{10}) in the thermodynamic limit $N\to \infty$ and obtains for
the free energy per spin:
\begin{equation}\label{13}
\frac{f}{N} \sim  \left\{
\begin{array}{lll}
&m^2+m^{\mu-1}-  mH,  & 2 < \mu <3 \, , \\
& m^2+ m^2 \ln \frac{1}{m} -  mH,  & \mu =3\, , \\
& \tau m^2+m^{\mu-1}- mH,  & 3 < \mu < 5\, , \\
& \tau m^2+m^4\ln\frac{1}{m}- mH,  & \mu = 5\, , \\
&  \tau m^2+ m^4-  mH,  & \mu>5\, .
\end{array}
\right.
\end{equation}
Here, $m$ is the order parameter and $\tau$  is the reduced
temperature.
A similar free energy  describes the critical behavior of the
standard IM ($\sigma_i=\pm 1$)  on an annealed scale-free network,
with  decay exponent $\lambda$  in that case \cite{Krasnytska16}
playing the role of $\mu$ in the current  one.
 The system is ordered at any finite temperature when $\mu\leq 3$ and has a second order phase transition when $\mu>3$.
All universal characteristics of the transition are $\mu$-dependent
when $3<\mu<5$ and the $\mu>5$ region is mean-field like. At $\mu=5$
logarithmic corrections feature
\cite{Dorogovtsev08,Lee09,Bianconi12,Krasnytska16,exponents},
governed by exponents which adhere to the usual scaling relations
\cite{logs}. These and leading exponents are listed in the third row
of Table~\ref{tab1} and discussed in further detail in that context.

For {\bf{Erd\"os-R\'enyi graphs}} one substitutes $p_{ij}=c$  into
(\ref{8}). This delivers a similar partition function (\ref{7})  to
that  of the complete graph  up to  renormalized interaction  so
that critical behaviors of both models are essentially equivalent.

{\bf For annealed scale-free networks,}
with $p_{ij}$ given by (\ref{5}),  the thermodynamic limit $N\to
\infty$ (i.e. with small $p_{ij}$) applied to (\ref{8}) gives
$d_{ij}\sim p_{ij}\beta J\mathcal{S}_i\mathcal{S}_j$. The
Stratonovich-Hubbard transformation  delivers the  trace over spins
in (\ref{7})  and a partition function that has  unary dependency on
the random variables  $f(k_i\mathcal{S}_i)$. It is convenient to
pass from the summation over nodes $i$ to summation over random
variables $k_i$, $\mathcal{S}_i$
$$
\sum_{i=1}^N f(k_i\mathcal{S}_i)=N\sum_{k=k_{\rm min}}^{k_{\rm
max}}\sum_{\mathcal{S}=\mathcal{S}_{\rm min}}^{\mathcal{S}_{\rm
max}}p(k)q(\mathcal{S})f(k, \mathcal{S})\, .$$ Considering {the}
variables $k$ and $\mathcal{S}$  {as} continuous and taking the
thermodynamic limit $N\to \infty$, we  put $k_{\rm
max}=\mathcal{S}_{\rm max}\to \infty$ and  we  choose the lower
bounds  $k_{\rm min}=\mathcal{S}_{\rm min} = 2$ \cite{Krasnytska20}.
 It is straightforward to see that the partition function  ${\cal Z}(\{\mathcal{S}\},\{k\})$ is independent of the random variables $k$ and $\mathcal{S}$ and is {\it self-averaging}.
Indeed,  when both distribution functions $p(k)$, $q(\mathcal{S})$
attain power-law forms (\ref{2}), (\ref{3}) we  obtain
\begin{equation} \label{14}
{ \cal Z} = \int_{0}^{+\infty} e^{\frac{-\overline{k}
T}{2N}x^2} \Big
[e^{I^{+}_{\lambda,\mu}(\sqrt{\frac{x}{N}})}+e^{I^{-}_{\lambda,\mu}(\sqrt{\frac{x}{N}})} \Big ]
\,dx\, ,
\end{equation}
and
\begin{equation}\label{15}
I^{\pm}_{\lambda,\mu}(\varepsilon)=N \Big [c_\lambda c_\mu \Big ( \frac{\varepsilon}{2}\Big )^{\lambda+\mu-2}I_{\lambda,\mu}(\varepsilon)
\pm\frac{\overline{\mathcal{S}^2}\, \overline{k}}{4T} \varepsilon^2 H\Big]
\end{equation}
 where
\begin{equation}\label{16}
I_{\lambda,\mu}(\varepsilon)=
\int_{\varepsilon}^{\infty}\int_{\varepsilon}^{\infty}
\varphi(k,\mathcal{S}) \, d\mathcal{S}dk \,
\end{equation}
with
\begin{equation}\label{17}
\varphi(k,\mathcal{S})=\frac{1}{k^\lambda \mathcal{S}^\mu}\ln
\cosh\Big( k\mathcal{S} \Big),
\end{equation}
and the lower integration bound $\varepsilon = 2 \sqrt{x/N}$ tends to zero, when $N\rightarrow
\infty$.

\begin{figure}
\centerline{\includegraphics[width=8cm]{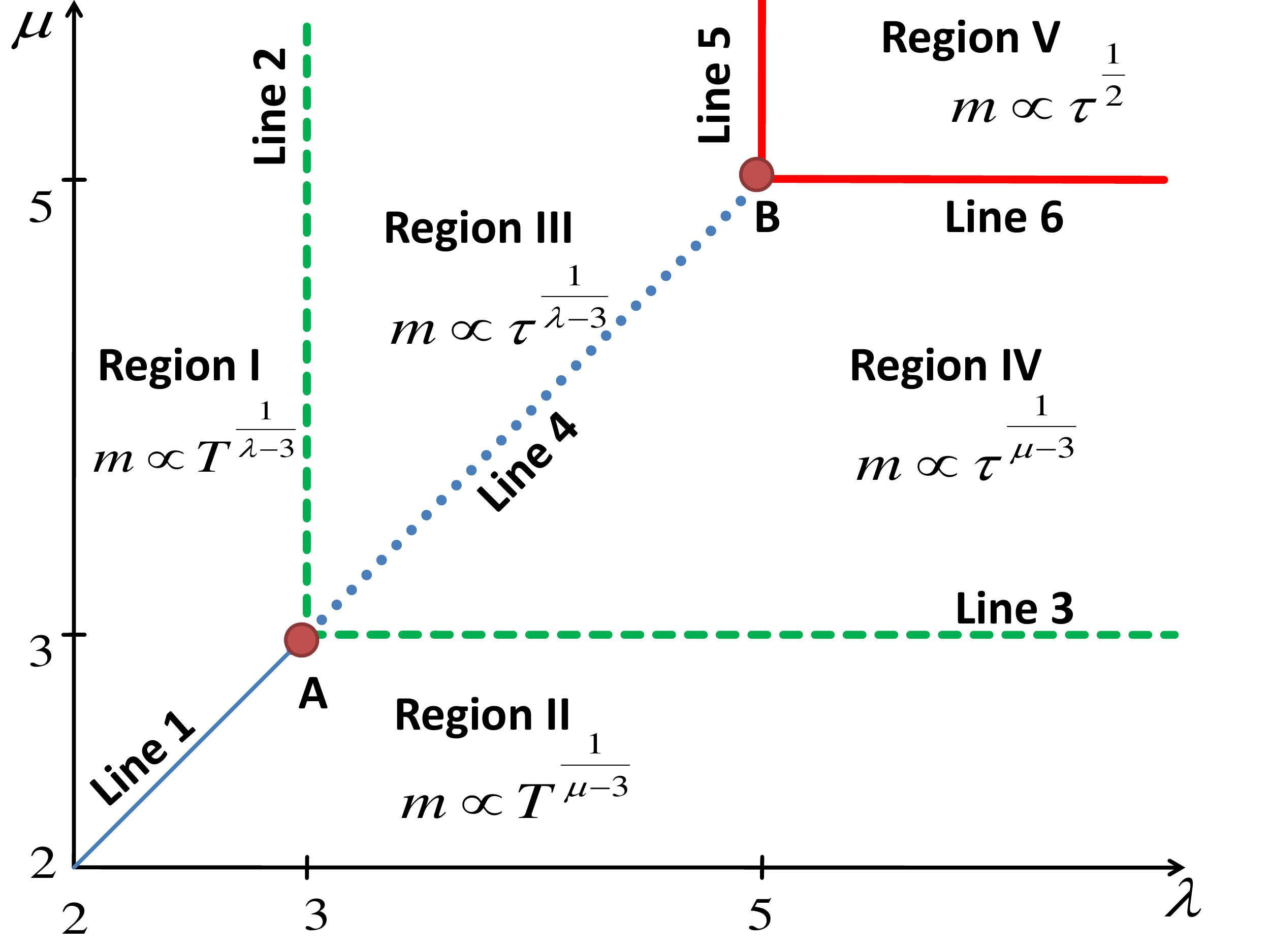}}
 \vspace*{8pt}
\caption{Phase diagram of the {IM}  with power-law distributed spin
strength (\ref{2}) on an annealed scale free network with node
degree distribution (\ref{3}). See Tables \ref{tab1}, \ref{tab2}.
  \label{fig1}}
\end{figure}

As in the  case of the model on the complete graph (\ref{12}),  the
partition function (\ref{14}) and hence  the of the free energy is
determined by the asymptotics of the integral (\ref{16}) as
$\varepsilon\to 0$. This is governed by the interplay of the decay
exponents $\lambda$ and $\mu$. In particular, for the diagonal case
$\lambda=\mu$ we get:
\begin{equation}\label{18}
I_{\lambda,\lambda}(\varepsilon)= a_\lambda + b_\lambda
\ln(\varepsilon) i_\lambda(\varepsilon)
 \, ,
\end{equation}
with
\begin{eqnarray}\label{A13}
i_\lambda (\varepsilon) \simeq  \left\{
\begin{array}{lll}
& O(\varepsilon^2),  & 2 < \lambda <3 \, , \\
&(\ln \varepsilon)^2/2 + O(\varepsilon^4),  & \lambda =3\, , \\
& {\frac{\varepsilon^{6-2\lambda}}{2(\lambda-3)^2}} + O(\varepsilon^2),  & 3 < \lambda < 5\, , \\
& \varepsilon^{-4}/8   + (\ln \varepsilon)^2/6 +O(\varepsilon^2),  & \lambda = 5\, , \\
& {\frac{\varepsilon^{6-2\lambda}}{2(\lambda-3)^2}} -
{\frac{\varepsilon^{10-2\lambda}}{12(\lambda-5)^2}}  +
O(\varepsilon^2),  & 5 < \lambda < 7\, .
\end{array}
\right.
\end{eqnarray}
The constants in (\ref{18}) can be readily evaluated and are
presented elsewhere \cite{Krasnytska20}. For the non-diagonal case
$\mu\neq \lambda$, due to the symmetry
$I_{\lambda,\mu}=I_{\mu,\lambda}$ it is enough to evaluate the
integral for $\lambda > \mu$. With  estimates  available from
\cite{Krasnytska20}, we  apply the steepest descent method to get
the exact solution for the partition function (\ref{14}). The
results that follow from the analysis of the free energy are
summarized in Fig. \ref{fig1} and \ Tables \ref{tab1}, \ref{tab2}.

Fig. \ref{fig1} presents the phase diagram of the model in the
$\lambda-\mu$ plane. Behavior is controlled by the parameter
with the smaller value. In Regions I and II, for which distributions
are fat-tailed  with $\lambda<3 $ or $\mu<3$, the system remains
ordered at any finite temperature $T$. There, the order parameter
decays with $T$ as a power law, $m\sim
T^{\frac{1}{\lambda-3}}$ or $m\sim T^{\frac{1}{\mu-3}}$
for $2<\lambda<3$ and $2<\mu<3$, correspondingly. Both asymptotics
coincide along Line 1  in the figure.

The decay is exponential $m\sim T e^{-T}$ along  Lines 2 and 3 where
$\lambda=3$ or $\mu=3$ as well as at point A where they coincide.
Second order phase transitions occur when both $\lambda,\mu>3$. In
Region III where $3<\lambda<5$ ($\mu>\lambda$) the critical
exponents  are $\lambda$-dependent and in Region IV where $3<\mu<5$
($\mu<\lambda$) they are $\mu$-dependent.

When $\lambda=5$ or $\mu=5$  logarithmic corrections to scaling
appear. For example, the order parameter in Lines~4-6 behaves as $m
\sim \tau^\beta |\ln \tau|^{-\hat{\beta}}.$ The values of the
logarithmic correction exponents in Lines 5 and 6 coincide with
those for the {IM}  on a scale-free network
\cite{Krasnytska16,exponents}. The additional richness of the phase
diagram is characterised, for example, by new type of logarithmic
corrections which emerge in Line 4 where $3<\lambda=\mu<5$ as well
as Point B where $\lambda=\mu=5$. All of them obey the scaling
relations for logarithmic corrections~\cite{logs}. Critical
exponents are summarized in Tables \ref{tab1} and \ref{tab2}.


Thus  introduction of variable spin or agent strengths to the  IM on
networks delivers rich new phase diagrams and universality classes
relevant to circumstances wherein interacting spins and agents carry
degrees of complexity over and above the binary features mostly
considered in physics and often rejected in other disciplines. On
the other hand, introducing random variables into the IM either in
form of annealed network or random spin strength makes the
interaction separable, as in  Mattis \cite{Mattis} or Hopfield
models \cite{Hopfield} used in description of spin glasses.

The model with power-law decaying random strength distribution
(\ref{2}) on the complete graph has similar critical behavior to the
standard {IM} on a scale-free network with random node degree
distribution (\ref{3}) with  decay exponents $\mu$ and $\lambda$
playing equivalent roles. This suggests that the level of complexity
introduced by allowing spin strengths to vary is on a par to the
level of complexity introduced by allowing  network architecture to
vary; i.e., individual strength  matter as much as connectivity.

When introduced to already rich annealed scale-free network the
complexity level is magnified yet more. Besides self-averaging,  it
is governed by the concurrence of  two parameters describing
different phenomena arising from  inherent interplay of two types of
randomness. The full phase diagram of the model (Fig. \ref{fig1}) is
symmetric under $\mu\leftrightarrow \lambda$ interchange,  critical
behavior governed by the smaller of the two parameters. Ordering,
critical behaviour and logarithmic corrections all feature.

The {IM}  itself taught us that interactions in physics play as important a role as spin properties, for without them we have no cooperative behavior or spontaneous magnetization. 
We have seen that spin strength and system architecture play similar counterbalancing roles and are tuned by the exponents $\mu$ and $\lambda$. 
{In sociophysics terms they suggest duality between strength of individual opinions and societal cohesion.
We hope that introduction of the former to the Ising model goes some way to addressing interdisciplinary objections that ``people are not atoms''~\cite{Stauffer}.
But we hope the variable-strength Ising model is suggestive of more than this too.
Region V of the phase diagram has high values of $\mu$ and $\lambda$ which may be interpreted as representing populism and weak society.
Regions I and II, on the other hand, represent empowerment of individuals and high social connectivity. 
Regions III and IV then represent the critical divide between these very different societies and suggests critical phenomena may play a role in further studies of the interplay between the individual and the collective.}

\section*{Acknowledgements} We thank Taras Krokhmalskii, Mykola Shpot, Reinhard
Folk, Tim Ellis, Volodymyr Tkachuk, Maxym Dudka and Yuri Kozitsky  for useful
discussions. It is our special pleasure and honour to thank
Camille No\^us from Laboratoire Cogitamus\footnote{http://www.cogitamus.fr/indexen.html} 
and doing so to declare our support of their acitivities.
Camille No\^us is a fictitious “collective individual” who embodies the academic community as a whole.
All new research, including that presented in this paper, is informed by the great body of human 
knowledge generated by colleagues past and present. In symbolizing this,
Camille is the antithesis of metricized “command and control” management of research.
The $\mu \leftrightarrow \lambda$ --type interplay between the individual and 
collective presented here is one that is frequently lost on academics and non-academics alike.
It is our hope that as this paper impacts the former, Camille impacts the latter.

\end{document}